\documentclass[aps,showpacs, a4paper,twoside,twocolumn]{revtex4}
\pdfoutput=1
\usepackage{graphics, subfigure, epsfig, color, moreverb, eufrak}
\graphicspath{{./figs/}}
\usepackage{ifpdf}
\ifpdf
\usepackage[pdftex,
	hypertexnames=false,
	bookmarksnumbered=true,
	colorlinks=true,
	pdfstartview=FitV,
	linkcolor=blue,
	citecolor=green,
	urlcolor=blue,
	hyperindex
            ]{hyperref}
\else
\fi

\newcommand{\qav}[1]{\left \langle #1 \right \rangle} 

\newcommand{\pa}[1]{\left ( #1 \right )}                
\newcommand{\reff}[1]{Fig.~\ref{#1}}
\newcommand{\refe}[1]{Eq.~(\ref{#1})}
\newcommand{\refs}[1]{Sec.~\ref{#1}}

\renewcommand{\vec}[1]{\mathbf{#1}}
\renewcommand{\Re}{\mathrm{Re }}

\newcommand{\mrm}[1]{\mathrm{#1}}

\newcommand{\be}{\begin{equation}}
\newcommand{\ee}{\end{equation}}
\newcommand{\bea}{\begin{eqnarray}}
\newcommand{\eea}{\end{eqnarray}}

\newcommand{\ev}[1]{\langle#1\rangle}
\newcommand{\ddt}{\frac{\partial}{\partial t}}

\newcommand{\mcal}[1]{{\mathcal{#1}}}
\newcommand{\drm}{{\mathrm{d}}}
\newcommand{\e}{\mathrm{e}}
\newcommand{\sinc}{\mathrm{sinc}}
\newcommand{\asinh}{\mathrm{asinh}}

\renewcommand{\vec}[1]{{\mathbf{#1}}}

\definecolor{light-gray}{gray}{.85}

\begin{document}
\title{Analytical Analysis of Single-Photon Correlations Emitted by Disordered
Semiconductor Heterostructures}
\author{P.~Bozsoki$^1$, W.~Hoyer$^2$, M.~Kira$^2$, I. Varga$^3$,
	P.~Thomas$^2$, S.W.~Koch$^2$, and H.~Schomerus$^1$ }
%
\affiliation{$^1$Department of Physics, Lancaster University, UK-LA1 4YB
Lancaster,
  United Kingdom}
\affiliation{$^2$Department of Physics and Material Sciences Center,
  Philipps-Universit\"at Marburg, D-35032 Marburg, Germany}
\affiliation{$^3$Elm\'eleti Fizika Tansz\'ek, Fizikai Int\'ezet, Budapesti
  M\H{u}szaki \'es Gazdas\'agtudom\'anyi Egyetem, Budafoki \'ut 8, H-1111
	Budapest, Hungary}
%
\date{\today}
%
\begin{abstract}
In a recent publication [Phys.\ Rev.\ Lett.~\textbf{97}, 227402 (2006), {\tt arXiv:cond-mat/0611411}],
it has been demonstrated numerically that a long-range disorder potential
in semiconductor quantum wells can be reconstructed reliably via
single-photon interferometry of spontaneously emitted light.

In the present paper, a simplified analytical model of independent
two-level systems is presented in order to study the reconstruction procedure
in more detail. With the help of this model, the measured photon correlations
can be calculated analytically and the influence of parameters such as the
disorder length scale, the wavelength of the used light, or the spotsize can
be investigated systematically. Furthermore, the relation between the proposed
angle-resolved single-photon correlations and the disorder potential can be
understood and the measured signal is expected to be closely related to the
characteristic strength and length scale of the disorder.
\end{abstract}
\pacs{78.55.-m, 42.50.-p, 71.35.-y, 78.30.Ly}
\maketitle
\section{Introduction}
\label{sec:intro}
%
Disorder in semiconductor heterostructures has strong influence on their
optoelectronic properties. Independently of whether disorder is introduced
by design or due to random interface roughness or compositional fluctuations
it can significantly alter the properties of the
heterostructure \cite{ot, es, ei, mtk,
Yayon:PRL,hardToMapDisorderToOptics-1, hardToMapDisorderToOptics-2}. Recent
publications show that long-range spatial disorder is an important topic
from both the technological and the fundamental point of view \cite{Yayon:PRL,
Bramwell:Nature, Gornyi:PRL, Langbein:PRL2002}.

In order to enhance our understanding of the role of disorder in these
systems, we have recently proposed an experimental scheme
\cite{Bozsoki:PRL06} which can be viewed as the Fourier analogue of nano-
or micro-luminescence \cite{wegener-1, wegener-2, nr4-1, nr4-2}. This
scheme is based on measuring angular correlations of spontaneously emitted
light and has been shown to give direct access to the spatial distribution
of the optically active electronic states and to the effect of disorder on
them. As our numerical investigations have shown the spatial distribution
can be recovered very reliably via a Fourier transformation of the
experimentally measurable photonic correlation \cite{Bozsoki:PRL06}.

In the present paper we present an extended analytical study using a
simplified model of uncoupled two-level systems. The additional
simplification introduced by neglecting the many-body interactions allows
us to deepen our understanding and present the fundamental principles
behind the suggested reconstruction procedure in a more transparent way.
One of the aims is to get a more intuitive understanding of the measured
correlation function, such that a basic understanding of the characteristic
parameters of the disorder landscape can be gained from the direct
measurement results without any reconstruction procedure.

The paper is structured as follows: We describe the experimental setup,
introduce the important photon-correlation functions, and outline the
reconstruction procedure in \refs{sec:Exp}. In \refs{sec:TLS} we present
general analytical results for a system of non-interacting two-level
systems where the disorder appears as a varying transition energy from site
to site. Finally, we choose a sinusoidal model potential instead of a truly
random disorder potential. For this special case, further analytical
results are derived in \refs{sec:model} before we conclude.

\section{Measuring correlations}
\label{sec:Exp}
%
Our suggested scheme is based on the angular photonic correlations of
spontaneously emitted light. Previous calculations have shown that not only
photon numbers $\ev{B^\dagger_{\vec{q}} B_{\vec{q}}}$, but also
single-photon correlations of the form $\ev{B^\dagger_{\vec{q}}
B_{\vec{q}'}}$ between photons of different modes ($\vec{q} \not=
\vec{q}'$) are building up when a semiconductor heterostructure reaches
quasi-equilibrium and spontaneously emits light \cite{PQE}. While the rate
of photons $\ddt \ev{B^\dagger_{\vec{q}} B_{\vec{q}}}$ is proportional to
the photoluminescence (PL) spectrum at energy $\omega_q = c|\vec{q}|$
and thus directly accessible to experiment \cite{PQE,Bozsoki:JLumi07}, a
clever setup must be used in order to measure the complex correlations. The
simplest possibility is to combine the light propagating along two
different directions in a common detector. Here, the light emitted into
directions $\vec{q}$ and $\vec{q}'$ is redirected with the help of mirrors
and through a beam splitter combined into the same detector. Consequently,
the detector detects the combined beam and the correct detector operator
for the description of the measurement process is given by
\be
d_{\vec{q},\vec{q}'} = B_{\vec{q}} + \e^{i \phi} B_{\vec{q}'},
\label{eq:detect_op}
\ee
i.e.\ a superposition of the photon operators of the two directions. The
variable phase $\phi$ is adjusted via the optical path length difference
between the two light beams. The detected signal is proportional to the
number operator corresponding to $d_{\vec{q},\vec{q}'}$ and thus given by
  \begin{eqnarray} \label{eq:AngularCorrelation}
	\ev{d^\dagger_{\vec{q},\vec{q}'} d_{\vec{q},\vec{q}'}}
	=
    \qav{ \pa{B_{\mathbf q}^\dag+\e^{-i \phi}B_{\mathbf q'}^\dag}
      \pa{ B_{\mathbf q}+\e^{i \phi}B_{\mathbf q'} }} \nonumber \\
=
      \qav{ B_{\mathbf q}^\dag  B_{\mathbf q}}
    + \qav{ B_{\mathbf q'}^\dag B_{\mathbf q'}}
    + 2\, \Re \qav{ \e^{i \phi} B_{\mathbf q}^\dag B_{\mathbf q'}}\,.
  \end{eqnarray}
The first two terms on the right hand side of \refe{eq:AngularCorrelation}
are nothing but the separate PL spectra emitted into the two different
directions while the interference term shows that also photon-correlations
between the two different emission directions can be made visible. As
previous results have shown, correlations can only build up if
$\vec{q}_{\|} = \vec{q}'_{\|}$ in the case of perfectly ordered
system~\cite{Hoyer:PRL04}. Here the parallel subscript denotes the
component of the wave vector along the quantum well (QW).

While the measured signal according to Eq.~(\ref{eq:AngularCorrelation})
still depends on a total of six components of $\vec{q}$ and $\vec{q}'$,
a few simplifications can be introduced according to the actual experiment;
firstly, a spectrometer is used in order to achieve a spectrally and
angularly resolved signal. The spectrometer is introduced after recombining
the two emission directions into one beam and before the detector.
Therefore, only the signal for equal $|\vec{q}| = |\vec{q}'|$ has to be
computed. Furthermore, our previous numerical investigations have shown
that the interference depends much stronger on the difference
$\vec{q}_{\|}-\vec{q}'_{\|}$ than on the average value such that we define
the angular correlation
  \begin{equation}\label{eq:U}
    U_{\hbar\omega}(\Delta \vec{q}_{\|})=\qav{ B_{\mathbf q}^\dag
B_{\mathbf q'}}
  \end{equation}
with the specific choice of $\vec{q}' = (\vec{q}_{\perp}, -\Delta\vec{q}_{\|}/2)$
and $\vec{q} = (\vec{q}_{\perp}, \Delta\vec{q}_{\|}/2)$ as the key observable.
The magnitude of the perpendicular component of $\vec{q}$ and $\vec{q}'$
has to be fixed for each choice of $\Delta\vec{q}_{\|}$ according to
$
\omega^2 = c^2 (|\vec{q}_{\perp}|^2 + |\Delta\vec{q}_{\|}|^2/4)\,.
$
Consequently, $U_{\hbar\omega}(\Delta \vec{q}_{\|})$ depends on only two
parameters, which is sufficient for a spectrally and angularly resolved scan.

\begin{figure}
  \centering
  \includegraphics[width=0.6\columnwidth, angle=-90]{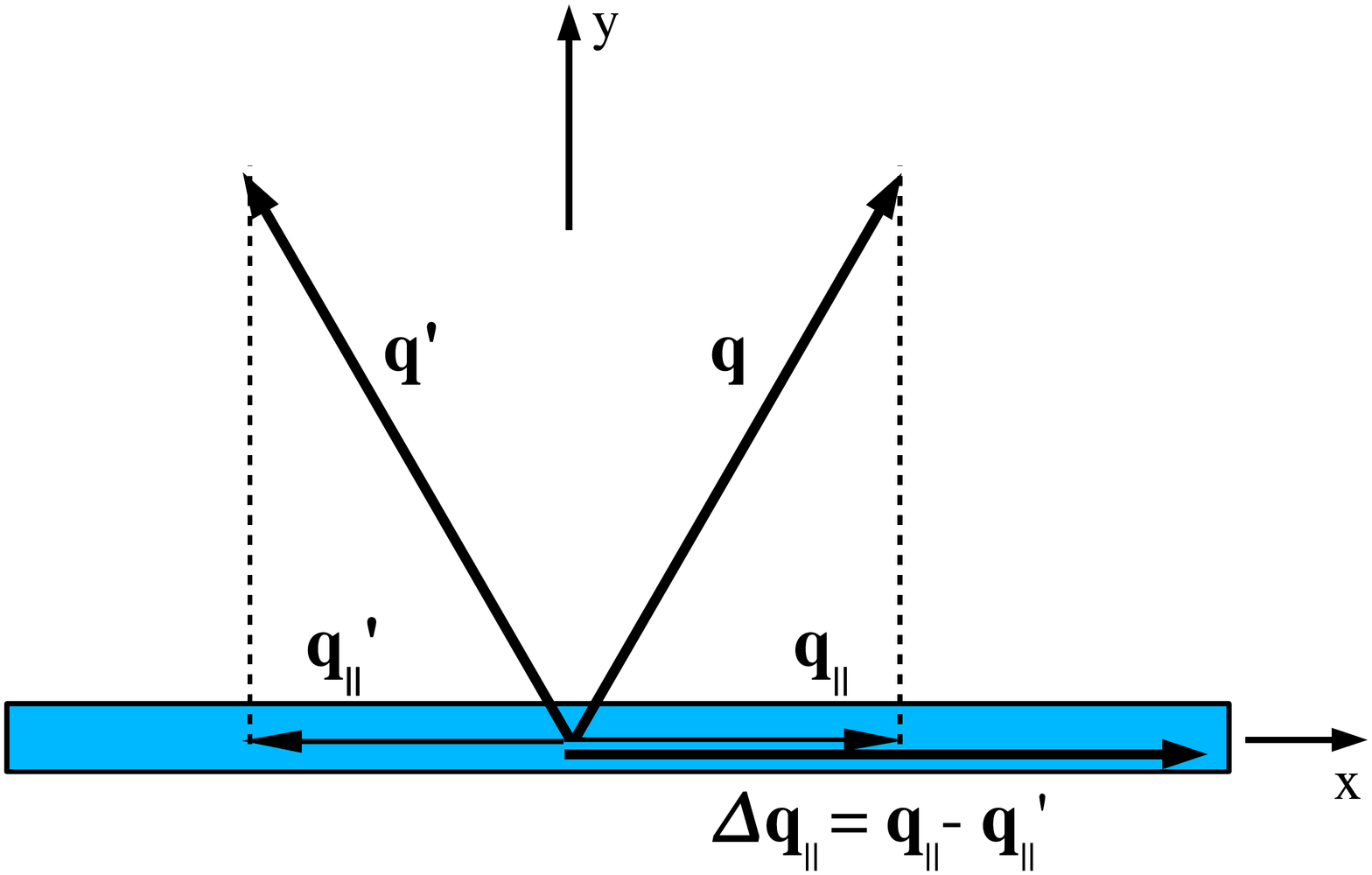}
  \caption{Schematic setup of a possible experimental arrangement for
	measuring photonic correlations.
\label{fig:expSetup}
}
\end{figure}

While the magnitude of $\vec{q}_{\perp}$ is determined by $\omega$ and
$\Delta\vec{q}_{\|}$, the sign can still be chosen differently. Our above
choice of identical signs for the perpendicular components of $\vec{q}$ and
$\vec{q}'$ represents measurement in e.g.\ reflection geometry as depicted
in \reff{fig:expSetup} which has the advantage that non-transparent
substrate can be used as both emission paths are on the same side of the
sample. An alternative possibility is to use light emitted into opposite
directions which corresponds to $\vec q'=-\vec q$. This setup was
successfully applied to detect angular photonic correlations in ordered
systems \cite{Hoyer:PRL04}.

The principle idea behind our scheme for the reconstruction of the disorder
potential exploits the fact that the strict momentum conservation along the
direction parallel to the QW is broken by a disorder potential. In contrast
to the ordered case where interference could only be observed for
$\Delta\vec{q}_{\|}=0$, we expect that non-zero correlations cannot only be
observed for vanishing $\Delta\vec{q}_{\|}$ but also for other values. Please
note that no such restriction applies to the direction perpendicular to the
QW, independently if it is perfectly ordered or disordered, because the
translational invariance is broken by the electronic confinement.

\subsection{Reconstruction procedure}
\label{sec:CorrelFunc}
%
For the sake of simplicity of the analytical calculations we consider a
one-dimensional quantum-wire in the remainder of the paper.  In this case,
the parallel component $q_\|$ and the spatial coordinate $x$ along the wire
are scalars. Apparently, all conclusions about momentum conservation hold
as for a QW. To simplify the notation we use $\Delta {q}$ instead of
$\Delta \vec{q}_{\|}$ in the rest of the present work. During the
calculations
we limit ourselves to two emission directions with both planes of emission
(spanned by the one-dimensional wire and the emission direction) being identical.

According to Eq.~(\ref{eq:AngularCorrelation}), the correlation function
from Eq.~(\ref{eq:U}) can be experimentally measured as an interference
contrast \cite{Bozsoki:PRL06}. By varying the optical path length $\phi$
in a controlled way, it is possible to extract both real and imaginary
part of $U_{\hbar\omega}(\Delta{q})$ such that it is legitimate to study
the full complex value of $U_{\hbar\omega}(\Delta{q})$.

By Fourier transforming the measured signal according to
\begin{equation}\label{eq:PosEmitters}
U_{\hbar\omega}(x)
	=
\frac{1}{2\pi}\int\limits_{-\Delta q_\mrm{max}}^{\Delta q_\mrm{max}}
U_{\hbar\omega}\pa{\Delta q} e^{-i \Delta q x} \,\drm{}\Delta q
\end{equation}
one can obtain information on the spatial distribution of emitting sources
of energy $\hbar \omega$. Here $\Delta q_\mrm{max}$ is determined by the
optical cone; since our scheme is a far-field method, the maximum value of
$\Delta{q}$ accessible in an experiment is obtained for oblique incidence
and given by approximately $\Delta q_\mrm{max}=2\omega_0/c=4\pi/\lambda_0$,
where $c$ is the speed of light in vacuum and $\lambda_0$ and $\omega_0$
are the wavelength and frequency of the average emission frequency.

For a fixed value of $\omega$, Eq.~(\ref{eq:PosEmitters}) can exhibit
multiple peaks at all those positions where the band gap corresponds to
$\hbar \omega$~\cite{Bozsoki:PRL06}. Thus, $U_{\hbar\omega}(x)$ can be
viewed as a probability distribution of the local contributions to the
emission spectrum. For a fixed value of $x$, on the other hand,
$U_{\hbar\omega}(x)$ has only a single peak at the local transition energy
such that the average energy can be defined as the weighted mean of the
transition energies via
  \begin{equation} \label{eq:potRecontruction}
      U(x) = {\int\limits_{E_1}^{E_2}
      \hbar\omega U_{\hbar\omega}(x) \,\drm{}\hbar\omega } /
{\int\limits_{E_1}^{E_2} U_{\hbar\omega}(x)  \,\drm{}\hbar\omega } \,.
  \end{equation}
Here, $E_1$ and $E_2$ are the limits of the energy scan carried out in the
experiment. $U(x)$ is called the \emph{reconstructed disorder
potential} \cite{Bozsoki:PRL06}.

It is important to realize that the reconstruction procedure can only
work within certain limits. Naturally, no disorder varying on length
scales larger than the spot size can be detected. Furthermore, as with
every far-field method, we are limited by the wavelength of the emission
and the reconstructed disorder potential can only be determined if it
varies on a typical length scale larger than the wavelenght of light.
In the present paper we demonstrate this scheme on the basis of a
theoretically
calculated $U_{\hbar\omega} \pa{\Delta q}$ for an ensemble of two-level
systems.

\section{Correlations of Two-Level Systems}
\label{sec:TLS}
%
As starting point we consider an ensemble of uncoupled two-level systems
as our model system. All two-level systems are spaced equally along $x$ with
lattice constant $a$ such that the $j^\mathrm{th}$ two-level system is positioned
at the lattice site $R_j = a j$. By using an equation-of-motion approach and
evaluating the Heisenberg equation of motion \cite{Bozsoki:PRL06}, we have
shown previously that the correlation function can be expressed as
  \begin{eqnarray}\label{eq:SLE-TLS}
    U_{\hbar\omega}(\Delta q) &=& \frac{2}{\hbar} |\mcal{F}_{\omega}|^2
   \sum_j \frac{\gamma\, S_j e^{i\Delta q R_j} } {\pa{E_j-\hbar\omega}^2 +
\gamma^2}  \,.
   \label{pum}
  \end{eqnarray}
Here, the prefactor contains the  matrix element $\mcal{F}_{\omega}$ of the
light-matter coupling including the microscopic interband dipole matrix
element, the so-called vacuum field amplitude, and the mode strength. The
remaining term is a sum over sites with homogeneously broadened peaks
centered at the site transition energies $E_j$ multiplied by the PL sources
$S_j=f_j^e f_j^h$, where $f_j^e$ and $f_j^h$ are the electron and hole occupation
probabilities of the respective site. The phase factor is a consequence of
the different emission directions and is the key ingredient for observing
the angle-resolved interference effects.

Taking the continuum limit $a \rightarrow 0$ with constant PL source
density ${S_j}/{a} \rightarrow S(x)$ and a continuously changing transition
energy $E(x)$, we can rewrite \refe{eq:SLE-TLS} as
  \begin{equation}\label{eq:corrFuncTLS}
    U_{\hbar\omega} \pa{\Delta q}
	=
\frac{2}{\hbar }
|\mcal{F}_{\omega}|^2
\int \frac{\gamma e^{i\Delta q x}S(x)}{\pa{E(x)-\hbar\omega}^2 +
\gamma^2}\,\drm{x} .
  \end{equation}
In the ordered case, all transition energies $E(x) = E_0$ are equal such
that the angular correlation is proportional to a single broadened
Lorentzian multiplied by the Fourier transform of the spot, $\int
\,e^{i\Delta q x}S(x) \,\drm{x}$.

\subsection{Spatial information}
\label{sec:SpatialInformation}
%
According to Eq.~(\ref{eq:PosEmitters}), the Fourier transform of the
correlation function gives some spatial information about the position
of emitters of a certain frequency $\omega$. Starting with the result
of Eq.~(\ref{eq:corrFuncTLS}), we obtain the Fourier transform
\bea
U_{\hbar\omega}(x)
& = &
\frac{2 |\mcal{F}_{\omega}|^2 \,\Delta{q}_\mrm{max} }{\hbar }
  \int
g_{\gamma}(E(y)-\hbar\omega) \,
\nonumber\\
&&\qquad\quad
S(y)\, \sinc(\Delta q_\mrm{max} (y-x))
\,\drm{y}
\label{eq:Ux}
\eea
where we introduced the broadened $\delta$-function
\be
g_{\gamma}(E) = \frac{1}{\pi} \frac{\gamma}{E^2 + \gamma^2}
\ee
and used the sine-cardinal $\sinc(x) = \sin(x)/x$.
The broadening of $g_{\gamma}$ is caused by the homogeneous line width
$\gamma$ and results in a finite energy resolution. But also
$\Delta{q}_{\mrm{max}} \,\sinc(\Delta{q}_{\mrm{max}} \,x)$ approaches a
Dirac delta function for large values of $\Delta{q}_{\mrm{max}}$. In that
case, the broadening of the sinc function for finite values of
$\Delta{q}_{\mrm{max}}$ reflects the fact that no spatial information can
be obtained for variations of the site energies on length scales below the
wavelength $\lambda_0=4\pi/\Delta{q}_{\mrm{max}}$.

\subsection{Dimensionless parameters}
\label{sec:dimless}
%
As one can anticipate from Eq.~(\ref{eq:Ux}), the possibility to obtain
precise spatial information strongly depends on the ratio $\gamma/W$
between homogeneous $\gamma$ and the disorder strength and on the ratios
$\lambda_0/L$ and $\mcal{S}/L$ between the observed wavelength or the spot
size $\mcal{S}$ and the typical disorder length scale $L$.

In order to obtain scaled equations with respect to the characteristic
parameters $W$ and $L$ of the disorder potential, we introduce $u=x/L$ and
$Q=L\Delta{q}$ as new spatial and momentum coordinates and
$\Omega=\hbar\omega/W$ and $\alpha=\gamma/W$ as the rescaled frequency and
homogeneous line width. The maximum value of $u$ is determined by the
spotsize to be $|u|<u_{max} =\mcal{S}/2 L$ while $Q$ is confined according
to $|Q|<Q_{max}=\frac{4\pi L}{\lambda_0}$. The dimensionless transition
energy is denoted $\tilde{E}(x) = E(x)/W$. With the help of these new
variables, we can rewrite Eqs.~(\ref{eq:SLE-TLS}) and~(\ref{eq:Ux}) as
\bea
\tilde{U}_{\Omega}(Q)
  & = &
\int g_{\alpha}(\tilde{E}(u)-\Omega) \,e^{i Q u} \, S(u) \, \drm{u} \,,
\label{eq:UQ_tls}
\\
\tilde{U}_{\Omega}(u)
& = &
\frac{1}{ \pi}
\frac{Q_\mrm{max}}{L}
 \int
g_{\alpha}(\tilde{E}(v)-\Omega) \, S(v) \nonumber \\
 && \hspace{1.5cm} \sinc(Q_\mrm{max} (v-u))
\,\drm{v}\,,
\label{eq:Uu_tls}
\eea
where we have eliminated the prefactors by the definition $U_{\hbar\omega}
= 2 \pi L\,|\mcal{F}_{\omega}|^2 /(\hbar W)  \tilde{U}_{\Omega}$ and
$g_{\alpha}(\tilde{E}) = \pi^{-1} \alpha/((\tilde{E}^2+\alpha^2) = W
g_\gamma(E)$. By virtue of the definition of $\tilde{E}$, its fluctuation
is on the order of unity and the energy resolution can be directly
estimated from the magnitude of $\alpha$. In the limit of strong disorder
($W\gg \gamma$) corresponding to $\alpha \rightarrow 0$ the broadened
$g_\alpha$ becomes a sharp Dirac delta function and we obtain
\small
\bea
\tilde{U}_{\Omega}(Q) \biggr|_{W \gg \gamma}
  & = &
\sum_n e^{i Q u_n} \, S(u_n)
\left(
	\frac{\drm \tilde{E}}{\drm{u}}\biggr|_{u=u_n(\Omega)}
\right)^{-1}
\\
\tilde{U}_{\Omega}(u) \biggr|_{W \gg \gamma}
& = &
\frac{1}{ \pi}
\frac{Q_\mrm{max}}{L}
\sum_n
\sinc(Q_\mrm{max} (u_n-u))  \nonumber \\
&& \hspace{.7cm} S(u_n)
\left(
	\frac{\drm \tilde{E}}{\drm{u}}\biggr|_{u=u_n(\Omega)}
\right)^{-1}\,,
\eea
\normalsize
where in both equations the sum over $n$ runs over all roots $u_n =
u_n(\Omega)$ of the argument of the delta function. These roots are defined
via $\tilde{E}(u_n(\Omega))=\Omega $. Obviously, only positions $u_n$
within the spot can contribute to the correlations, as is guaranteed by the
factor $S(u_n)$ which vanishes outside the excitation spot.

Another interesting limit is obtained for a disorder length scale well
above the wavelength of the emitted light, but still below the spotsize. In
that limit $Q_{\mrm{max}}\gg 1$ such that the sine cardinal function can be
treated as a sharp delta function and we find
\bea
\tilde{U}_{\Omega}(u) \biggr|_{L \gg \lambda_0}
& = &
\frac{1}{ \pi}
\frac{Q_\mrm{max}}{L}  g_{\alpha}(\tilde{E}(u)-\Omega) \, S(u) \,.\,\,
\eea
In that case, the spatial resolution is very good even for relatively large
values of $\gamma$ since for fixed $u=x/L$ the correlation function
$U_{\Omega}(u) $ has its maximum at the transition energy $\hbar\omega =
E(x)$. According to Eq.~(\ref{eq:potRecontruction}) the reconstruced
potential follows exactly the actual energetic variation $E(x)$.

\section{Model Potential}
\label{sec:model}
While the results of the previous section are valid for a periodic
arrangement of non-interacting two-level systems with arbitrary disorder
(i.e.\ arbitrary variation of the transition energies), we will introduce
additional approximations in the present section in order to simplify the
equations even further. In particular, we study the interpretation of
$\tilde{U}_{\Omega}(Q)$, i.e., the direct measurement signal. That way, we
can see what information on the disorder potential can be gained  from the
measurement without performing any reconstruction.

Firstly, we assume an infinitely large spot and a homogeneous luminescence
source $S(x) \equiv S_0$. Secondly, we consider a very specific variation
of transition energies in the form $\tilde{E}(u) = E_0+ \sin(2\pi u)$.
While a real disorder potential in general is more randomly fluctuating and
might even contain several characteristic length scales, our more regular
model potential has a perfectly well defined length scale.

With the above mentioned approximations, we can simplify
Eq.~(\ref{eq:UQ_tls}) for $\Omega=E_0$ and get
\bea
\tilde{U}_{\Omega=E_0}(Q)
  & = &
\int g_{\alpha}(\sin(2\pi u)) \,e^{i Q u} \, \drm{u} \,,
\label{eq:UQ_model}
\eea
After rewriting the broadened Lorentzian function as
\small
\be
g_{\alpha}(\sin(2\pi u))
=
\frac{i}{2\pi} \left(
      \frac{1}{\sin(2 \pi u) + i\alpha}-
      \frac{1}{\sin(2 \pi u) - i \alpha}\right)\,,
\label{eq:sinDecompose}
\ee
\normalsize
we can solve the integral in Eq.~(\ref{eq:UQ_model}) with the help of
Cauchy's theorem where the contour of the integration has to be closed in
the upper (lower) complex plane for positive (negative) sign of $Q$,
respectively.

The poles of the two contributions of Eq.~(\ref{eq:sinDecompose}) are
given by the roots of the energy denominators. The roots of the first
denominator are defined by the equation
\be
\sin(2 \pi u) = - i\alpha\,.
\label{eq:defpoles}
\ee
Introducing the quantity
\be
y_0 = \asinh(\alpha) = \ln\pa{\alpha+\sqrt{1+\alpha^2}}\,,
\ee
we can express the solutions of Eq.~(\ref{eq:defpoles}) as
\be
u^{(m)}_{1,-} = -\frac{i}{2\pi} y_0 + m\,,
\quad
u^{(m)}_{1,+} = +\frac{i}{2\pi} y_0 + m + \frac12\,,
\ee
for integer values of $m$. The subscripts $+$ and $-$ signify whether the
pole lies in the upper or lower complex plane. Similarly, the poles of the
second contribution of Eq.~(\ref{eq:sinDecompose}) are given by
\be
u^{(m)}_{2,-} = -\frac{i}{2\pi} y_0 + m + \frac12\,,
\quad
u^{(m)}_{2,+} = +\frac{i}{2\pi} y_0 + m\,.
\ee

The residues of the integrand of Eq.~(\ref{eq:UQ_model}) are given by
%
\small
\be
\mrm{Res}\left[ \frac{i}{2\pi} \frac{e^{i Q u} }{\sin(2 \pi u ) + i\alpha}; u^{(m)}_{1,\pm} \right]
 =
\pm \frac{1}{2\pi i} \, \frac{1}{2\pi}\frac{e^{i Q u^{(m)}_{1,\pm}} }{\cosh(y_0 )}\,,
\ee
\be
\mrm{Res}\left[ \frac{i}{2\pi} \frac{e^{i Q u} }{\sin(2 \pi u ) - i\alpha}; u^{(m)}_{2,\pm} \right]
 =
\mp\frac{1}{2\pi i} \, \frac{1}{2\pi}\frac{e^{i Q u^{(m)}_{2,\pm}} }{\cosh(y_0 )}\,,
\ee
\normalsize
%
such that after inserting the explicit solutions for $u^{(m)}_{1/2,\pm}$ we obtain
\bea
\tilde{U}_{\Omega=E_0}(Q)
  & = &
\frac{1}{2\pi}
\frac{\e^{-\frac{|Q| y_0}{2\pi}}}{\cosh(y_0)}
\sum_{n=-\infty}^{\infty} \e^{i Q \,n/2} \,.
\label{eq:UQ_solution}
\eea
%
Employing the identity
  \begin{equation}
    \sum_{n=-\infty}^{\infty}e^{iQ\,n/2}= 4\pi\sum_{m=-\infty}^{\infty}
    \delta(Q-4\pi m)\,
  \end{equation}
%
we obtain the final result
\bea
\tilde{U}_{\Omega=E_0}(Q)
  & = &
4\,
\frac{g^{-\frac{|Q|}{2\pi}}}{g + g^{-1}}
\,
\sum_{m=-\infty}^{\infty} \delta(Q-4\pi m)\,.
\label{eq:UQ_final}
\eea
%
Here we have introduced
%
\begin{equation}
 g=\alpha+\sqrt{1+\alpha^2}
\end{equation}
%
and used the definition of $y_0$ as well as the relations
%
\bea
\e^{-\frac{|Q| y_0}{2\pi}} & = & \left(\cosh(y_0) +
\sinh(y_0)\right)^{-\frac{|Q|}{2\pi}} = g^{-\frac{|Q|}{2\pi}} \,,
\\
\cosh(y_0) &=& \sqrt{1+\alpha^2} = \frac{1}{2} (g+g^{-1})\,.
\eea
%

Our result, Eq.~(\ref{eq:UQ_final}), nicely demonstrates that for the
sinusoidal model potential the expected angular correlation consists of
regularly spaced delta functions multiplied with an envelop function which
exponentially decays with $|Q|$. The decay rate with respect to $|Q|$ is
determined by $\ln(g)$ and thus a function of the ratio $\alpha=\gamma/W$.
The different peaks are located at integer multiples of $4\pi$
corresponding to a spacing of $4\pi/L$ for the unscaled variable
$\Delta{q}$.

\begin{figure}[tp]
  \centering
  \includegraphics[width=0.8\columnwidth, angle=0]{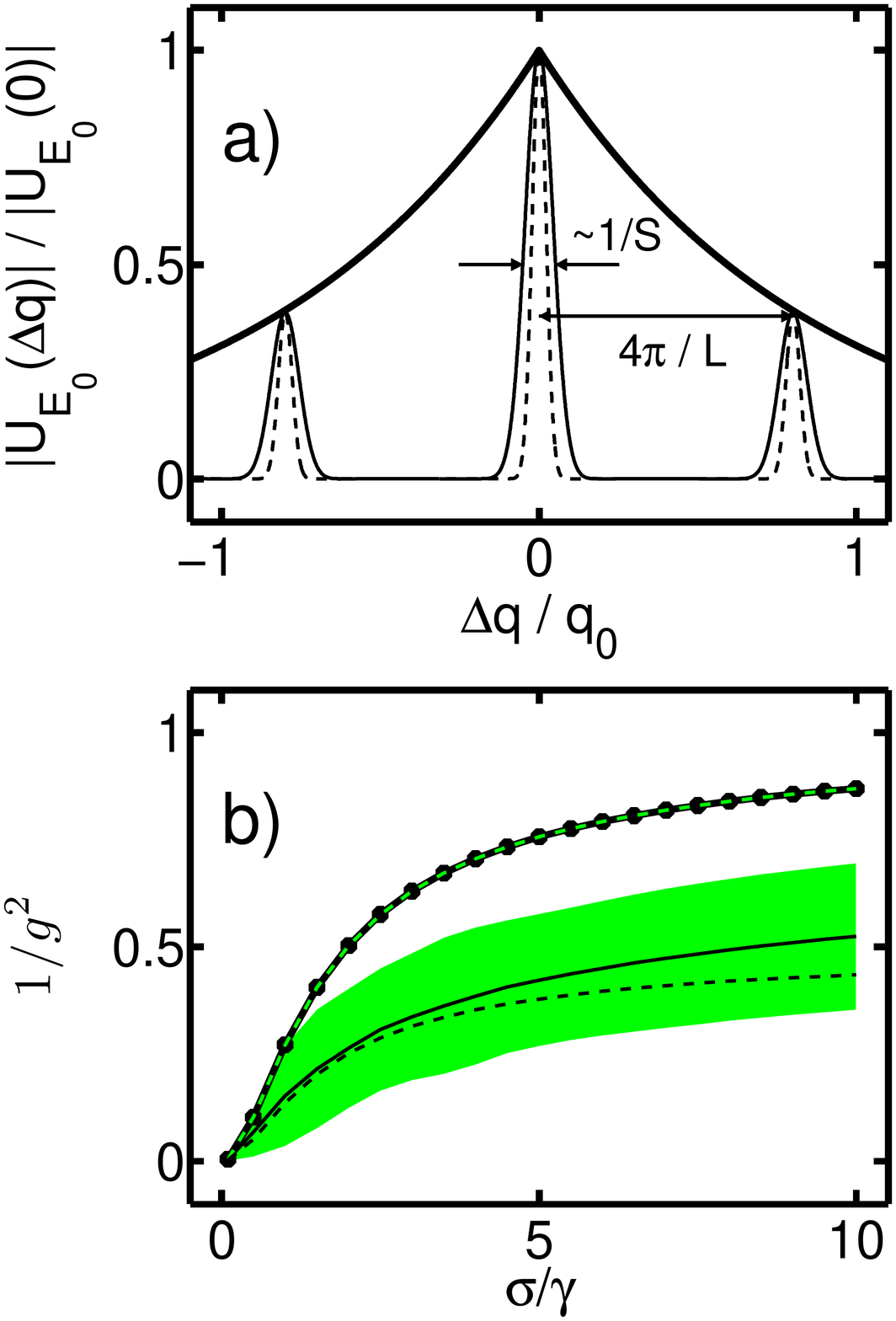}
  \caption{%
    a) Correlation function $U_{\hbar\omega= E_0}(\Delta q)$ as function
    of $\Delta{q}/q_0$ for a sinusoidal potential and different spotsizes.
    Comparison between the analytically computed envelope function (thick
	solid line) and the numerical result for a spotsize of $S=10$\,$\mu$m
	(thin solid line) and $S=20$\,$\mu$m (dashed line). Furthermore
	indicated are the peakwidth proportional to the inverse spotsize
	and the peak distance inversely proportaion to the disorder length scale.
     Numerical parameters are
    $\gamma=1$ meV, $W/\gamma=20$, $L=1$ $\mu$m.\\
    b)  The height of the second maximum of $U_{\hbar\omega=
      E_0}(\Delta q)$ as a function of the variance of the disorder
    potential. Thick line: sinusoidal potential, analytical
    result. Dots: numerical results for arbitrary spot size. The
    dashed line gives the thick line divided by 2. For random
    potentials of various length scales $L$ the resulting curves are
    situated in the shaded area. The thin line is a guide for the
    eye.
\label{fig:TLS-sin}}
\end{figure}

In order to confirm our results and study the influence of a finite spot
size, we display the envelope function $4 g^{-\frac{|Q|}{2\pi}}/(g+g^{-1})$
together with the results of two numerical computations for finite spot
size in \reff{fig:TLS-sin}a. Here, we have normalized the result to a
maximum value of 1 and plotted the correlation as a function of
$\Delta{q}/q_0$ where the wavenumber $q_0$ corresponds to an assumed
emission wavelength of $\lambda_0 = 800$\,nm. The distance between
neighbouring sites has been chosen $a=5$\,nm.

We notice that our predictions are confirmed by the numerical simulations
and that furthermore the finite spotsize does not influence the envelope
function. While the width of the peaks is inversely proportional to the
spotsize, both their height and their position do not depend on $\mcal{S}$.
In a real experiment, we can thus expect that the distance between the
central peak and the first neighbouring peak is a signature of the longest
disorder length scale involved, while the ratio between the peak heights of
first and second peak, given by
\be
\frac{\tilde{U}_{\Omega=E_0}(4 \pi)}{\tilde{U}_{\Omega=E_0}(0)} =g^{-2}\,,
\ee
can provide a measure of the ratio between homogeneous $\gamma$ and the
disorder fluctuation $W$. For known homogeneous broadening, one can thus
extract information on the energetic spread of the transition energies.

A good measure of the energy fluctuations for very different kinds of
disorder potentials is given by the standard deviation $\sigma$. For a
general random potential used in a numerical computation, the standard
deviation is obtained via \be
\sigma^2 = \frac{1}{N} \sum_{j} (E_j - \bar{E})^2
\ee
as a sum over the lattice sites, where $\bar{E}$ is the average energy and
$N$ is the total number of sites. For our sinusoidal model potential in the
continuum limit, we must use the integral form
\be
\sigma^2 = \frac{1}{L} \int_0^L \sin^2(2\pi x/L)\,\drm{x} = W^2/2\,.
\ee
The result shows that the standard deviation in that case is simply
proportional to the amplitude of the sinusoidal function. For a comparison
between the analytical model system and numerical solutions with random
disorder potentials, a systematic study of the ratio $g^{-2}$ as function
of $\sigma/\gamma$ is shown in \reff{fig:TLS-sin}b. As a consequence of the
random disorder potential, also the peak ratios for different realizations
are different. The shaded area denotes the region in which all numerical
results of a large number of simulations have been found. The analytical
result (thick solid line) is found to be roughly twice as large over the
whole range of $\sigma$ values. Analytical tests with a superposition of
more than one sine function have shown that the analytical result for such
a model also decreases towards the numerically observed region.

\section{Conclusions}
\label{sec:Conclusions}
%
In conclusion, the presented analytical derivation of angular
photonic correlations supports the results of previously published numerical
investigations~\cite{Bozsoki:PRL06} and helps to understand the underlying
principles of the suggested reconstruction scheme.

We have derived the angular correlation function for an ensemble of
uncoupled two-level systems. In that model, the correlation function has
been found to be given by the Fourier transform of the excitation spot for
a perfectly ordered system while the presence of disorder shows up in a
deviation from this shape. The spatial distribution
of emitters can be recovered via a Fourier transformation. For a sinusoidal
model potential analytical formulas have been derived for the dependency of
the primary experimentally detectable signal on the characteristic strength
and length scale of the disorder potential.\\

\acknowledgments
The authors are grateful to H.\ Stolz, K. Maschke, and T. Meier for valuable
discussions. P.\ B.\ and H.\ S.\ gratefully acknowledge the financial support
by the European Commission, Marie Curie Excellence Grant
MEXT-CT-2005-023778 (Nanoelectrophotonics). In Marburg, this work has been
supported by the Optodynamics Center of the Philipps-University Marburg
and by the Deutsche Forschungsgemeinschaft through the Quantum Optics in
Semiconductors Research Group. I.V. thanks for financial support from
OTKA (Hungarian Research Fund) under Contracts No.\ T042981 and No.\ T46303.


\end{document}